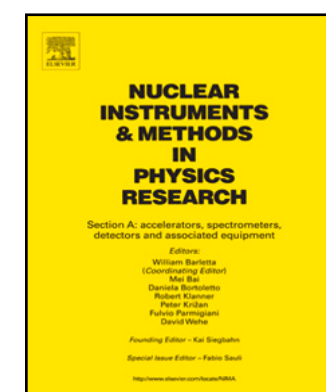

Full Length Article

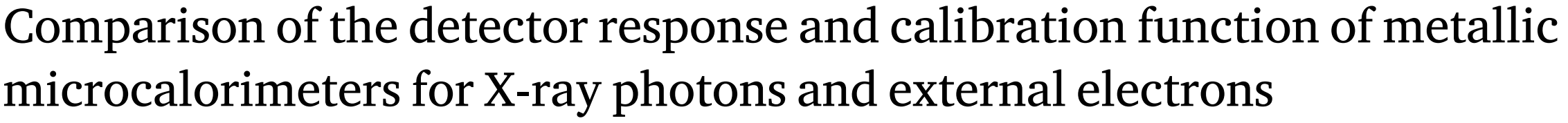

# Comparison of the detector response and calibration function of metallic microcalorimeters for X-ray photons and external electrons

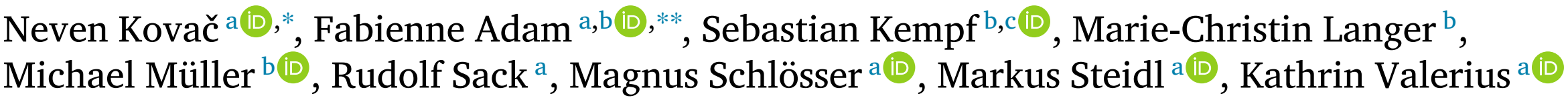

Neven Kovač [a,*], Fabienne Adam [a,b,**], Sebastian Kempf [b,c], Marie-Christin Langer [b], Michael Müller [b], Rudolf Sack [a], Magnus Schlösser [a], Markus Steidl [a], Kathrin Valerius [a]

[a] Institute for Astroparticle Physics (IAP) - Karlsruhe Institute of Technology (KIT), Hermann-von-Helmholtz-Platz 1, 76344, Eggenstein-Leopoldshafen, Germany

[b] Institute of Micro- and Nanoelectronic Systems (IMS) - Karlsruhe Institute of Technology (KIT), Hertzstraße 16, 76187, Karlsruhe, Germany

[c] Institute for Data Processing and Electronics (IPE) - Karlsruhe Institute of Technology (KIT), Hermann-von-Helmholtz-Platz 1, 76344, Eggenstein-Leopoldshafen, Germany

## ARTICLE INFO

## ABSTRACT

Metallic microcalorimeters (MMCs) are cryogenic single-particle detectors that rely on a calorimetric detection principle. Due to their only thermally limited theoretical energy resolution, close-to-ideal linear detector response, fast signal rise time and the potential for 100 % quantum efficiency, MMCs outperform conventional detectors by several orders of magnitude. These attributes make them particularly interesting for a broad spectrum of applications, including a next-generation neutrino mass experiment based on the measurement of the tritium beta-decay spectrum, with an objective of achieving a sensitivity surpassing that of the pioneering KATRIN experiment. However, although MMCs have been used in measurements of photons and heavy ions with great success, no information is currently available on the interaction between MMCs and external light charged particles such as electrons. This work aims to provide such missing information and to demonstrate that MMC-based detectors are suitable for high-resolution spectroscopy of external electron sources. Particularly, we present the first-ever measurements of external electrons using a metallic microcalorimeter, comprehensively discuss the characteristics of the signal shape and the calibration function and give a direct comparison between well-defined conversion electron and X-ray photon signals from the same $^{83}$Rb/$^{83m}$Kr source.

## 1. Introduction

The discovery of neutrino oscillations at the end of last millennium provided compelling evidence that neutrinos are massive particles [1,2]. Due to their copious production in the early Universe, neutrinos played a key role in structure formation, as well as in the evolution and expansion of the Universe [3]. However, oscillation experiments cannot provide any information on the absolute scale of the neutrino mass and a different, direct approach is required. Today, the measurement of the absolute scale of the neutrino mass in a model-independent way is one of the most important and pressing questions in contemporary physics.

The only model-independent experimental approach is based on kinematic analysis of weak decay processes, most notably the $\beta$ decay of tritium [4]. Here, the effect of non-zero neutrino mass manifests itself as a distortion and a shift of the end-point region of the measured electron energy spectrum from the tritium decay at 18.6 keV [5].

Currently, by far the most constraining limits on the neutrino mass obtained from direct measurements come from the KATRIN (**KA**rlsruhe **TRI**tium **N**eutrino) experiment hosted at the Tritium Laboratory Karlsruhe (TLK) [6].

The KATRIN experiment consists of a 70 m long beamline, designed to probe the electron anti-neutrino mass by high-resolution and high-statistics measurements of the electron spectrum, produced in the beta-decay of a gaseous molecular tritium source. The working principle of the KATRIN experiment is based on the MAC-E (Magnetic Adiabatic Collimation with Electrostatic) filter technology [7–9], which analyses the energy of individual electrons by acting as a high-pass energy filter. By counting the number of electrons that pass the filter at each predefined retardation energy using a multi-pixel silicon *p-i-n*-diode array, an integral spectrum can be reconstructed. Recently, an upper limit on the electron anti-neutrino mass was determined to be $m_{\bar{\nu}_e} \leq 0.45\,\mathrm{eV}/c^2$ at 90% CL [10]. The final sensitivity of the KATRIN experiment is about $0.3\,\mathrm{eV}/c^2$ [6], after 1000 days of measurement, and is expected to be reached at the end of 2025.

* Corresponding author.
** Corresponding author at: Institute for Astroparticle Physics (IAP) - Karlsruhe Institute of Technology (KIT), Hermann-von-Helmholtz-Platz 1, 76344, Eggenstein-Leopoldshafen, Germany.
*E-mail addresses:* neven.kovac@kit.edu (N. Kovač), fabienne.adam@kit.edu (F. Adam).

The current technology employed by KATRIN does not allow for significant scaling to reach beyond the final sensitivity. Therefore, future neutrino-mass experiments necessitate the development of next-generation technology in order to reach the inverted mass regime, and possibly below that. To this end, KATRIN++ has been proposed as a next-generation neutrino-mass experiment, with the goal to identify and develop new scalable technology by making use of the existing infrastructure of KATRIN. Two primary concepts have been identified as key for the success of the future experiment: (1) an ultra-high-resolution differential detection method and (2) a high luminosity atomic tritium source.

Currently, KATRIN uses a molecular tritium source in which spectral broadening originating from molecular final states limits the effective resolution of the measurement to about $1\,\mathrm{eV}$ FWHM [11]. In a future experiment an atomic source needs to be employed in order to mitigate this limitation.

Furthermore, switching from an integral measurement mode, currently employed at KATRIN, to a differential one, will allow for a much more efficient collection of statistics, a substantial reduction of background and an increase in the energy resolution of the experiment. New detector technology will need to be able to resolve the energy of individual electrons, with an energy resolution (FWHM) better than $1\,\mathrm{eV}$. One option for such detectors are cryogenic quantum sensors such as metallic microcalorimeters (MMCs) which, at the time of writing of this article, provide a record energy resolution of $\Delta E_{\mathrm{FWHM}} = 1.25\,\mathrm{eV}$ for $5.9\,\mathrm{keV}$ soft and tender X-ray photons [12,13], with future prospects of reaching sub-eV energy resolution. This fact makes MMCs a highly appealing candidate for a detector in a future neutrino mass experiment using tritium. MMCs have already been employed for example in the framework of the ECHo experiment aiming to determine the electron neutrino mass by performing a calorimetric measurement of the electron capture spectrum of a $^{163}$Ho source implanted in the MMC absorber [14]. However, so far MMCs have never been used for measurements of electrons emitted by an external source, to the best of our knowledge. For this reason, systematic effects on the energy resolution due to the interaction between an MMC and external electrons such as backscattering, sputtering and creation of Frenkel pairs need to be experimentally studied and very well understood. In the work at hand, we present the first-ever measurements of external electrons using a metallic microcalorimeter. We comprehensively discuss the characteristics of the signal shape and the calibration function and give a direct comparison to the signals from X-ray photons.

## 2. Magnetic microcalorimeters (MMCs)

Magnetic microcalorimeters (MMCs) are cryogenic single-particle detectors that rely on a calorimetric detection principle [15,16]. They consist, as schematically shown in Fig. 1, of an application-specific particle absorber and a paramagnetic temperature sensor that are thermally strongly coupled to each other. Moreover, the absorber is weakly coupled to a heat bath with temperature $T < 50$ mK, allowing the detector to thermalize at a well-defined baseline temperature in the absence of an energy input. The temperature sensor is exposed to a weak external magnetic field of a few mT to induce a temperature-dependent sensor magnetization. In this configuration, a temperature change resulting from an energy deposition in the absorber causes a decrease in the magnetization of the paramagnetic sensor, which can be measured with highest precision by means of a superconducting flux transformer and a direct-current superconducting quantum interference device (dc-SQUID). The transformer is formed by a superconducting pick-up coil magnetically coupled to the temperature sensor as well as the input coil of a current-sensing SQUID.

To fully stop the impinging particle, the absorber has to be made out of a material featuring a high stopping power for the particle type to be detected. Gold has been the dominant material choice for photon detection due to it being a radiopure noble metal, that is easy to deposit by microfabrication techniques, and has no metastable states. In addition, gold possesses a good thermal conductivity and a high atomic number, which is favorable since the absorption of photons occurs primarily via the photoelectric effect [17,18]. The downside of gold is its comparably large heat capacity, which spoils the energy resolution. As paramagnetic sensor material, gold or silver doped with a few hundred ppm of the rare earth metal erbium has established as the state of the art over the last decades [15,19].

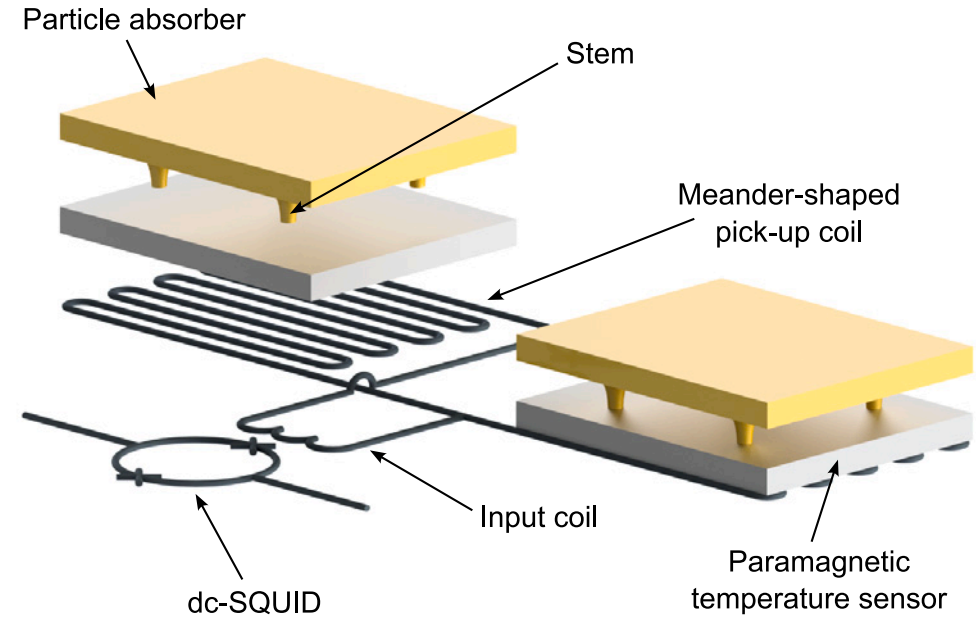

**Fig. 1.** Schematic sketch of the state of the art detector geometry with two meander-shaped pick-up coils connected in parallel to the input coil of a current-sensing dc-SQUID. Absorber and planar temperature sensor are connected to each other via stems, which reduce energy loss to the substrate due to athermal phonon escape.

Various detector geometries exist for realizing an MMC [15,16]. The current design of choice is based on a planar, superconducting meander-shaped pick-up coil [15,20]. Here, as schematically depicted in Fig. 1, the overall pick-up coil consists of two planar meander-shaped coils that are gradiometrically connected to the input coil of the current-sensing dc-SQUID. On top of each coil, a temperature sensor and a particle absorber are placed. In addition to detecting the magnetization change within the paramagnetic temperature sensor, the meander-shaped coils are used to generate the magnetic field by carrying a persistent current. Due to the gradiometric geometry of the pick-up coils, the magnetic field will have same magnitude but different sign, leading to opposite polarization of the two sensors. In this way, the signals from the two pixels will have different polarity, too, allowing them to be distinguished from each other.

## 3. Measurement set-up

All measurements presented in this paper were performed in a commercial dilution refrigerator with a base temperature below $7\,\mathrm{mK}$. We used a custom detector set-up made out of copper to guarantee a reliable thermalization of the detectors. Fig. 2 shows a close-up of the setup that mainly consists of a massive copper bar, a detector platform, a $^{83}$Rb/$^{83\mathrm{m}}$Kr source, and a hollow aluminum cylinder that is mounted on a copper carrier plate and serves as superconducting shielding.

The detector platform consists of a massive copper plate that is equipped with two independent detector chips, each featuring four individual, two-pixel MMCs based on the meander-shaped geometry shown in Fig. 1. The overall setup hence comprises eight pixels in total. The detector chips were originally fabricated in the framework of the PrimA-LTD project [21] and were optimized for measuring the electron capture spectrum of $^{55}$Fe. Hence, each $175\,\mu\mathrm{m} \times 175\,\mu\mathrm{m}$-sized gold absorber measures $12\,\mu\mathrm{m}$ in thickness. Each MMC channel is read-out by one custom current-sensing dc-SQUID located on a separate first-stage SQUID chip placed directly next to the detector chip to minimize parasitic inductances induced by the wiring between the chips. For the measurement, the detector platform is covered by a microfabricated gold collimator with openings slightly smaller than the size of the particle absorbers to prevent charging effects by electrons hitting the substrate. The detector platform is mounted at the end of a 12.5 cm long copper bar close to the bottom of the surrounding aluminum cylinder.

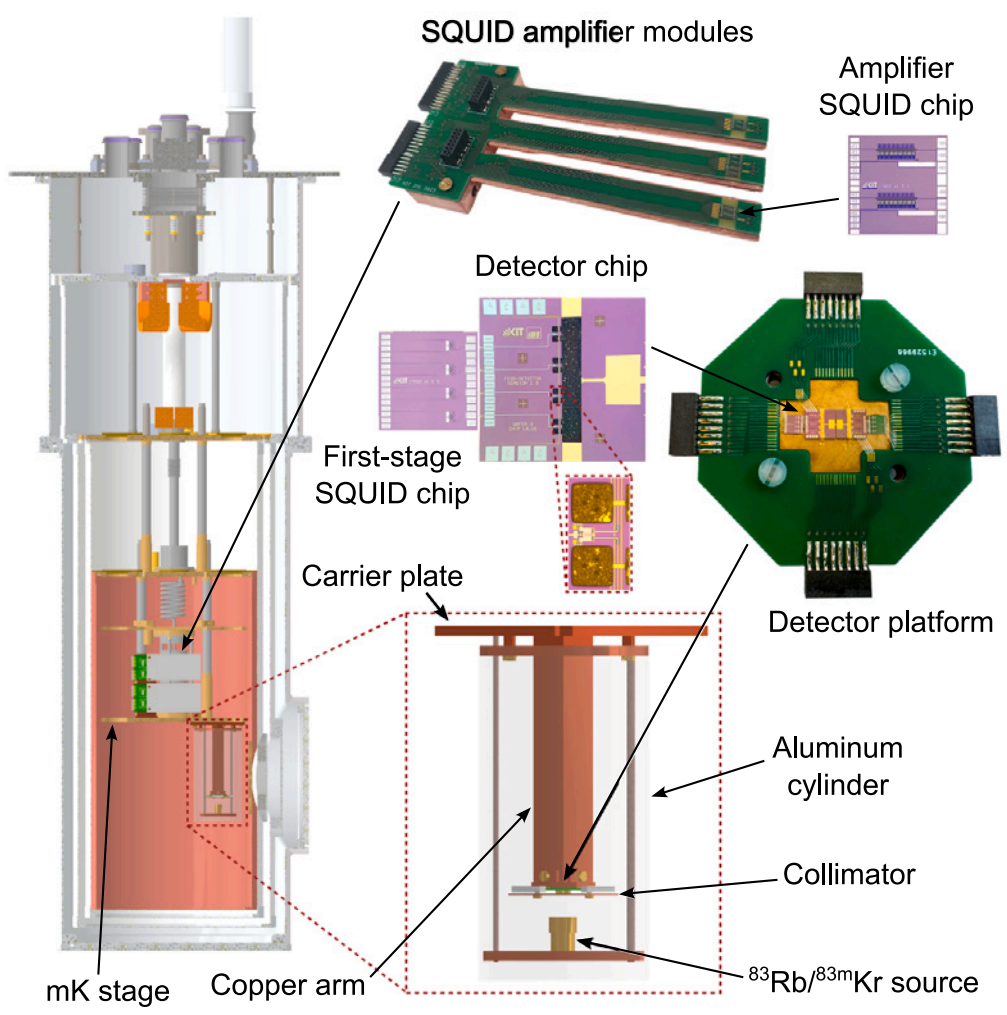

**Fig. 2.** Schematic of the cryogenic measurement setup. A detailed description of all components is given in Section 3.

Inside the aluminum cylinder, a $^{83}$Rb/$^{83m}$Kr source is mounted facing the detector platform. It is attached to the same copper plate carrying the massive cooper bar hosting the detector platform but with two separate threaded rods to minimize thermal contact between the source and the detector. The electron source is based on the metastable $^{83m}$Kr isomer which is produced in the decay of $^{83}$Rb with high efficiency of 74(5)% and a half-life of 86.2(1) days [22]. The $^{83}$Rb atoms are implanted into a (10×10) mm$^2$-sized highly-oriented pyrolytic graphite substrate (HOPG) at the Bonn Isotope Separator (BONIS) [23]. As a source holder, the HOPG substrate is embedded into a semi-open gold-plated copper cylinder. With a half-life of 1.8 h, $^{83m}$Kr decays via a cascade of two gamma transitions of 9 keV and 32.2 keV energy which are highly converted [24]. Therefore, the resulting spectrum consists of multiple mono-energetic conversion electron lines in the range between 7.5 keV and 32.1 keV as well as several X-ray lines, the most notable ones being the $KX_\alpha$ line at 12.6 keV and the $KX_\beta$ line at 14.1 keV [24].

The first-stage dc-SQUIDs reading out the MMCs were connected to custom $N$-dc-SQUID series arrays acting as a low-temperature amplifier stage for the first-stage dc-SQUID signals. Each dc-SQUID series array was connected to commercially available room temperature electronics. The amplifier SQUIDs were assembled in custom SQUID array modules that are surrounded by superconducting as well as soft-magnetic shielding. These modules are separately attached to the mK stage of the cryostat to avoid heat dissipation in the vicinity of the detector module.

## 4. Results

In order to characterize the response of MMCs to external electrons, we measured the $^{83m}$Kr decay spectrum[1] with an average event rate per pixel of about 0.1 s$^{-1}$. As $^{83m}$Kr provides both conversion electrons and X-ray photons, direct comparison of the detector response upon photon and electron impact can be obtained with a single measurement. We acquired ≥ 100 pulses for each selected $^{83m}$Kr line and averaged the recorded signals for each line. Averaging improves the signal-to-noise ratio (SNR), allowing the detector response to be more easily compared and, in particular, to see potential spurious effects. In addition, we removed pulses which deviate noticeably in shape and size.[2]

[1] Total activity of the source at the time of measurement was about 36 kBq.

**Table 1**
Energy and number of averaged pulses for the $^{83m}$Kr lines selected for the analysis. The energy values of the respective lines are taken from [24]. $KX_{\alpha2}$ line is not used in the analysis, and the information is only included here for completeness.

| Line | Particle type | Energy (keV) | # Pulses |
|---|---|---|---|
| $L_{2,3}-32$ | $e^-$ | 30.4516 | 363 |
| $K-32$ | $e^-$ | 17.8242 | 142 |
| $KX_{\beta1,\beta3}$ | X-ray | 14.1102 | 1177 |
| $KX_{\alpha1}$ | X-ray | 12.648 | 5487 |
| $KX_{\alpha2}$ | X-ray | 12.595 | 3719 |
| $M_1-9$ | $e^-$ | 9.1129 | 105 |
| $L_1-9$ | $e^-$ | 7.4811 | 524 |

We selected the following $^{83m}$Kr spectral lines for the analysis: higher energy $L_{2,3}-32$ and $K-32$ conversion electron lines, $KX_{\alpha1}$ and $KX_{\beta1,\beta3}$ X-ray lines and lower energy $M_1-9$ and $L_1-9$ conversion electron lines. Naming convention for the $^{83m}$Kr lines follows the convention used in [24], with a few exceptions: $L_{2,3}-32$ is a doublet consisting of $L_2$ and $L_3$ peaks from 32.3 keV gamma transition and $KX_{\beta1,\beta3}$ is an X-ray doublet consisting of $KX_{\beta1}$ and $KX_{\beta3}$ peaks, which cannot be resolved at the current energy resolution of the measurement. The number of selected pulses as well as the energy of each selected line are reported in Table 1. Fig. 3 shows the averaged signal pulses in linear and semi-logarithmic representation as well as a zoom into the signal rise.

### 4.1. Shape of the detector signal

For an ideal microcalorimeter that can be modeled by a canonical ensemble with two sub-systems representing the absorber and the thermometer, each signal pulse can be described by the sum of two exponentials representing the signal rise and decay [16]. However, as the finite system bandwidth of the SQUID read-out introduces a cutoff frequency $f_r$ (usually in the range of 1 MHz), the detector response function used to describe the signal induced magnetic flux change $\Delta\Phi$ in the first-stage SQUID needs to be modified, taking the form [25]:

$$\Delta\Phi(t) = A \cdot \left\{ \frac{\tau_1}{\tau_1 - \tau_r}\left(e^{-t/\tau_1} - e^{-t/\tau_r}\right) - \frac{\tau_0}{\tau_0 - \tau_r}\left(e^{-t/\tau_0} - e^{-t/\tau_r}\right) \right\} . \quad (1)$$

Here, $A$ is the amplitude of the pulse, $\tau_0$ is the signal rise time constant, $\tau_1$ is the signal decay (thermalization) time constant and $\tau_r$ is the characteristic time constant given by $\tau_r = 1/(2\pi f_r)$, with $f_r$ being the cutoff frequency of the readout system. In addition, depending on the complexity of the detector design, the detector response might show not a single but several decay constants, with one dominant exponential function which can be attributed to the thermalization of the detector with the heat bath via weak thermal link.

In order to compare the detector response to external electrons and X-ray photons, each signal pulse shown in Fig. 3 is fitted with the function from Eq. (1). For an ideal detector, we expect no difference in the time constants for signal from electrons and from photons. Fig. 4(a) displays an example of such a fit. Prior to fitting, each pulse is scaled to a unitary amplitude. The fit region is limited to $t \in [0, 0.5]$ ms, as in this region the contribution from additional exponential functions in the decay part is negligible. Fig. 4(b) shows the comparison of the fit results for the signal rise time (yellow) and signal decay time (blue).

[2] Removed pulses make up less than 0.1% of the total number of pulses, comprising primarily of noise signals, such as mobile phone signals, and pile-up events.

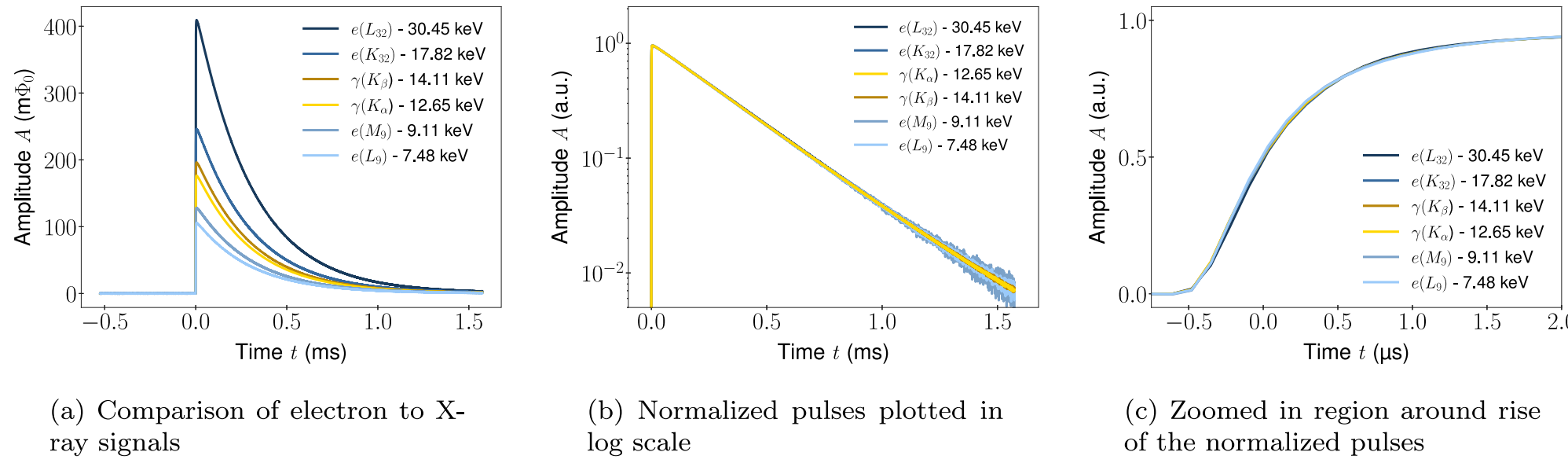

(a) Comparison of electron to X-ray signals

(b) Normalized pulses plotted in log scale

(c) Zoomed in region around rise of the normalized pulses

**Fig. 3.** MMC detector response to the energy input from photons (shades of yellow) and electrons (shades of blue) emitted by the $^{83}$Rb/$^{83m}$Kr source. We selected in total six $^{83m}$Kr spectral lines for the analysis: two from X-ray photons and four from conversion electrons. The plots show (a) averaged signal pulses, (b) normalized averaged signal pulses in logarithmic scale and (c) normalized averaged signals, zoomed in the region around the signal rise.

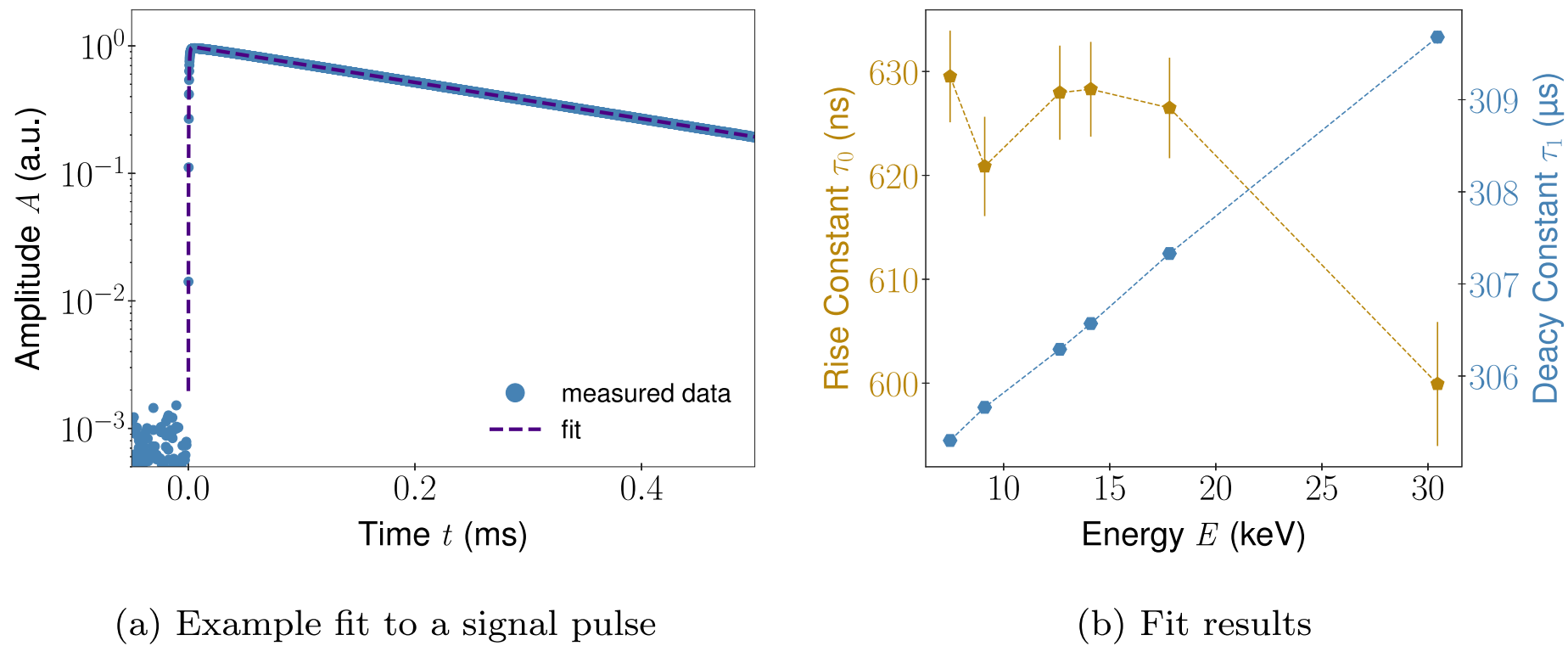

(a) Example fit to a signal pulse

(b) Fit results

**Fig. 4.** (a) Example fit to a single pulse using the fit function from Eq. (1). (b) Rise time constant $\tau_0$ and decay time constant $\tau_1$ versus line peak energy as determined from fitting Eq. (1) to the different average signals shown in Fig. 3. Note: Uncertainties on the determined decay times constants are too small to be visible in the plot.

The decay time constant weakly increases with increasing energy of the particle, which can be attributed to a small change in the heat capacity and thermal conductance associated with a larger temperature increase for larger energy inputs. For similar reasons, the rise time shows a slight temperature dependence. Overall, the agreement between the shape of the different averaged signal traces is excellent and indicates that there is no systematic deviation between signals resulting from electrons and from photons. The obtained fit parameters are consistent with expectations from the detector design, and does not show a dependence on the particle type.

### 4.2. Pulse shape consistency

An important check that the detector responds to photons and external electrons in the same way is to compare the difference between signal pulses of both particle types. For the comparison, we selected pulses from several krypton lines, namely: $KX_{\alpha 1}$ and $KX_{\beta 1,\beta 3}$ X-ray lines, $L_1{-}9$, $M_1{-}9$ and $K{-}32$ conversion electron lines. As these signals correspond to lines of different energies, each pulse is scaled to a unitary amplitude. Signals that were chosen for the comparison are those closest to one another in energy. This is important as small energy dependence in the pulse shape is expected due to change in thermodynamic properties of the detector, i.e. heat capacity and thermal conductance, for different energy inputs.[3]

In the data acquisition software used for the presented measurement, each signal pulse is acquired with a time window of about 2 ms, split into 16 384 discrete samples. Therefore, after subtraction of two pulses we obtain a set of 16 384 points representing difference between pulses in different pulse regions. If the pulse shapes associated with different signals are the same, a histogram constructed from the calculated differences between individual points of two pulses is expected to be Gaussian distributed, with the width depending on the level of detector noise. Non-Gaussian shape featuring additional peaks or kinks would indicate a difference in the shape of the pulses, i.e. in the signal rise parts and/or decay parts.

Fig. 5 shows three such plots, each with three histograms showing the pulse difference between two X-ray signals (blue), two electron signals (green) and finally the difference between an electron and an X-ray signal (yellow). As different parts of the pulse can differ for different particle types, comparison was performed in different regions of the pulse, notably around the rise part (0.1 ms region around maximum), around the rise part and first part of the decay (0.5 ms region around maximum) and taking most of the pulse into consideration (1.0 ms region around maximum). Fig. 5 shows the resulting histograms of differences; in (a) for the 0.1 ms region, in (b) for the 0.5 ms region, and finally, in (c) for the 1.0 ms region around the signal maximum. All the histograms show the expected Gaussian shape, which was confirmed by performing a Gaussian fit to the distributions, giving a mean $\mu$ consistent with 0 for all nine histograms. Standard deviation $\sigma$ for electron–gamma histograms (yellow) and gamma–gamma histograms (blue) is on the order of $10^{-3}$, while electron–electron histograms (green) feature a slightly larger $\sigma$ of $1.5 \times 10^{-3}$, which can be attributed to a smaller amplitude of the signals and therefore worse SNR. Values of the fit parameters are reported in Table 2, together with the reduced-$\chi^2$ of the respective fits. In addition, no visible structures appeared in the residuals, as shown in Appendix.

[3] These effects are expected to be small in the energy range of the krypton spectrum, but nonetheless we chose pulses from lines that are closest to each other in energy.

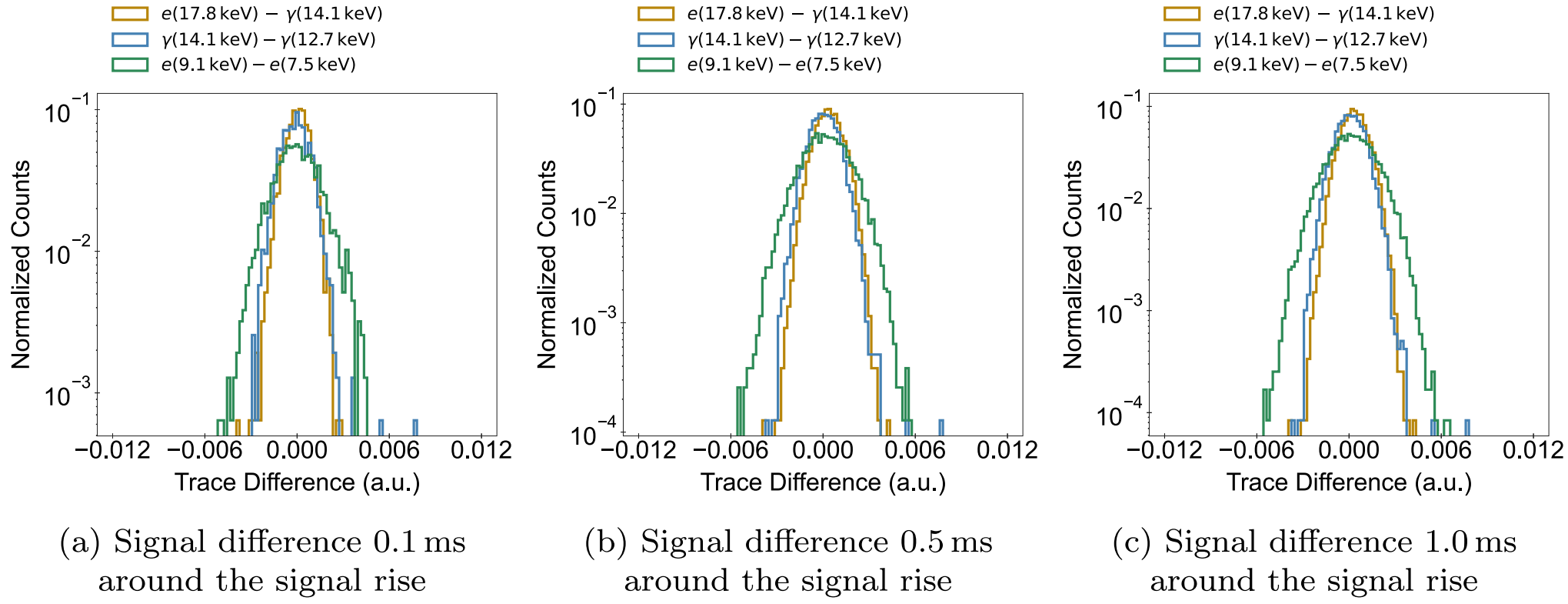

(a) Signal difference 0.1 ms around the signal rise

(b) Signal difference 0.5 ms around the signal rise

(c) Signal difference 1.0 ms around the signal rise

**Fig. 5.** Histograms of the differences between individual signal pulses: electron versus electron signal, photon versus photon signal and electron versus photon signal. Signals were chosen such that they are closest to one another in energy, and all histograms are normalized to 1. Individual panels correspond to the pulse difference (a) in the region of 0.1 ms around the signal rise, (b) in the region of 0.5 ms around the signal rise, and (c) in the region of 1.0 ms around the signal rise.

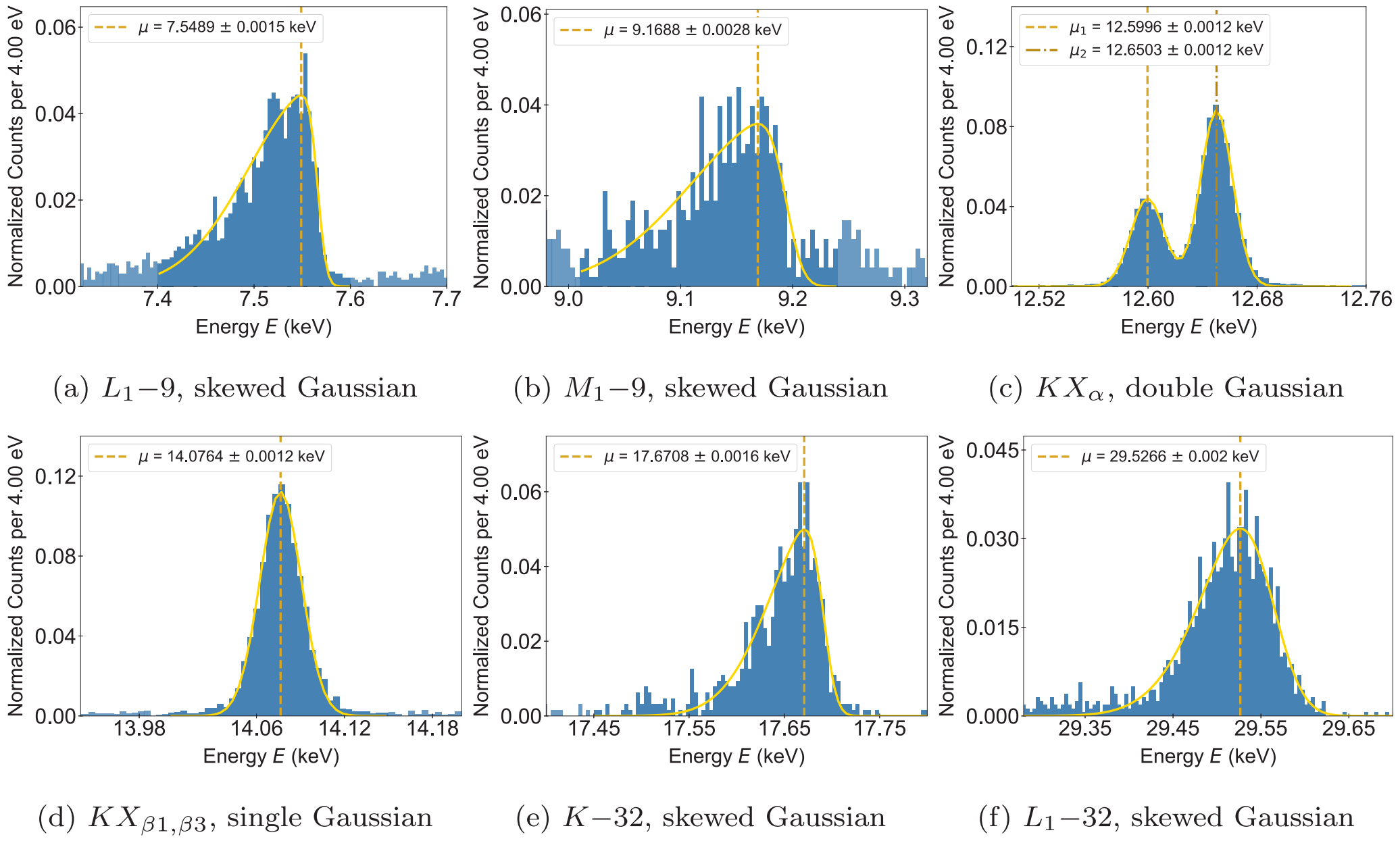

(a) $L_1-9$, skewed Gaussian

(b) $M_1-9$, skewed Gaussian

(c) $KX_\alpha$, double Gaussian

(d) $KX_{\beta1,\beta3}$, single Gaussian

(e) $K-32$, skewed Gaussian

(f) $L_1-32$, skewed Gaussian

**Fig. 6.** Fits to selected spectral lines of the $^{83m}$Kr spectrum: (a) fit to the $L_1-9$ conversion electron peak with a skewed Gaussian, (b) fit to $M_1-9$ conversion electron peak with a skewed Gaussian, (c) fit to the $KX_\alpha$ X-ray peak with a double Gaussian function, (d) fit to the $KX_{\beta1,\beta3}$ X-ray peak with a single Gaussian function, (e) fit to the $K-32$ conversion electron peak with a skewed Gaussian and (f) fit to the $L_1-32$ conversion electron peak with a skewed Gaussian. Obtained peak positions $\mu$ with their uncertainties are reported in the legend.

This detailed analysis shows that pulse shape is the same and independent of the particle type and the energy within a given energy window.

### 4.3. Energy calibration function

Another important piece of information is whether the non-linearity and the calibration function differ for external electron and X-ray photons. For MMCs, as for other cryogenic detectors, some non-linearity is expected, i.e., the measured energy is not linearly related to the energy of the incoming particle. This is due to the fact that the thermodynamic properties of the detector depend on temperature and slightly change with the energy of the incoming particle [15]. To estimate the non-linearity, we fit a polynomial function to the measured amplitudes of the six selected $^{83m}$Kr peaks and use the information on the true energy of the peaks, taken from [24]. To estimate the amplitude of individual peaks, each peak is fitted with an appropriate function; all electron peaks are fitted with a skewed Gaussian function which takes into account the effect of backscattering of electrons on the absorber of the detector.[4] On the other side, the $KX_\alpha$ doublet peak is fitted with a double Gaussian function due to the proximity of the $KX_{\alpha1}$ and $KX_{\alpha2}$ peaks (as reported in Table 1), and the $KX_{\beta1,\beta3}$ peak is fitted with a single Gaussian function.[5] The peak amplitude is then given by the position(s) of the maxima of the respective fitted Gaussians. Fig. 6 shows the measured histograms of the different peaks as well as the applied fit functions. The corresponding line positions are reported in the legend together with their respective uncertainty.

**Table 2**
Fit parameters, $\mu$ and $\sigma$, from the Gaussian fit to the histograms of pulse differences, shown in Fig. 5. Reduced-$\chi^2$ of the individual fits is also reported in the table and the residuals are shown in Appendix.

| Signal difference | Time window around signal rise | $\mu$ | $\sigma$ | $\chi^2$/ndf |
|---|---|---|---|---|
| $e^-$ - X-ray | 0.1 ms | 0.000 038 | 0.000 794 | 1.16 |
| X-ray - X-ray | 0.1 ms | −0.000 139 | 0.000 958 | 1.06 |
| $e^-$ - $e^-$ | 0.1 ms | −0.000 036 | 0.001 486 | 1.21 |
| $e^-$ - X-ray | 0.5 ms | 0.000 379 | 0.000 925 | 1.01 |
| X-ray - X-ray | 0.5 ms | 0.000 056 | 0.000 957 | 1.38 |
| $e^-$ - $e^-$ | 0.5 ms | 0.000 192 | 0.001 563 | 1.27 |
| $e^-$ - X-ray | 1.0 ms | 0.000 357 | 0.000 891 | 1.32 |
| X-ray - X-ray | 1.0 ms | 0.000 076 | 0.000 958 | 1.38 |
| $e^-$ - $e^-$ | 1.0 ms | 0.000 21 | 0.001 547 | 1.16 |

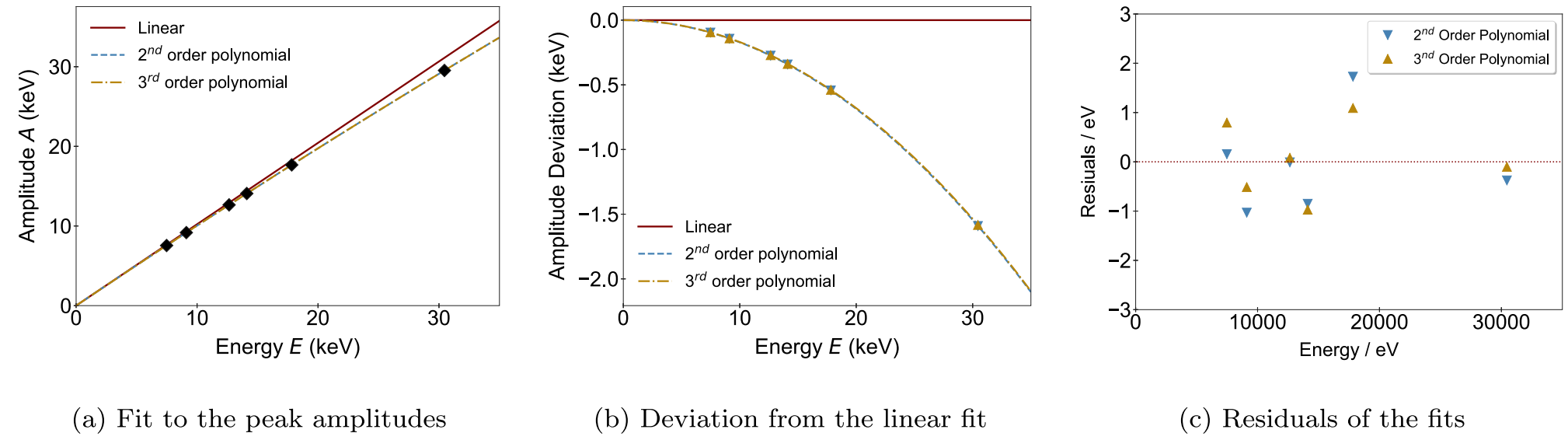

(a) Fit to the peak amplitudes (b) Deviation from the linear fit (c) Residuals of the fits

**Fig. 7.** Calibration function for the measured $^{83m}$Kr spectrum: (a) fit to the peak amplitudes with a 2nd and 3rd order polynomial functions, with the linear component shown as well, (b) deviation from the linear fit, i.e., non-linearity, (c) residuals of the data points from the fit function.
*Source:* Theoretical values for the peak energies taken from [24].

Using the obtained peak amplitudes together with the known energy of individual spectral lines, the calibration function can be obtained from a polynomial fit. The fit was performed with both second and third order polynomial functions, both of which are shown in Fig. 7(a). The use of second or third order polynomial functions is justified by the fact that for larger energy inputs, thermodynamical properties of the detector have to be taken into account [15].

In Fig. 7(b), the difference between these polynomial fits and the linear behavior is shown, from which we obtain a non-linearity of only about 5% within the entire 30 keV range. Additionally, Fig. 7(c) shows the residuals for both fit cases, all of which are within the $3\sigma$-uncertainty on the peak position. We can conclude that the second order polynomial fit is sufficient and all of the peaks, both from electron and from photon signal, are described with the same calibration function. The determined non-linearity of the detector,[6] being around 5% at 30 keV, is well within the expected range for a detector of such kind and design [15,21].

## 5. Conclusion and outlook

The main motivation of this work was to demonstrate that MMC detectors can be employed for the detection and spectroscopy of external light charged particles, i.e., electrons. The possibility to perform such measurements would allow MMCs to be used in the field of high-resolution electron spectroscopy. Such detectors could form an enabling technology for next-generation neutrino-mass experiments, aiming to probe sub-100 meV region of the neutrino mass and ultimately go beyond the inverted mass ordering regime. The results presented in this work stand, to the best of our knowledge, as the first-ever confirmation of feasibility to perform such MMC measurements, and demonstrate that there is no apparent difference in the detector response to energy deposits from external electrons versus X-ray photons. This implies that in the analysis signals from photons and external electrons do not need to be treated separately and the same analysis strategy developed over the course of previous years can be directly transferred from one use case to the other. Similar measurements have recently been conducted with Transition Edge Sensors (TESs), confirming the feasibility to use these detectors for high resolution electron spectroscopy as well [26].

The measured $^{83m}$Kr spectrum will be discussed in full details in the upcoming publication, after all the systematic effects, such as backscattering, have been fully investigated.

Looking towards the future, by building upon this work we plan to develop a quantum sensor demonstrator to further test and advance this technology for future experiments.

## CRediT authorship contribution statement

**Neven Kovač:** Writing – review & editing, Writing – original draft, Software, Methodology, Investigation, Formal analysis. **Fabienne Adam:** Writing – review & editing, Writing – original draft, Visualization, Software, Methodology, Investigation, Conceptualization. **Sebastian Kempf:** Writing – review & editing, Validation, Supervision, Resources, Project administration, Methodology, Investigation, Funding acquisition, Conceptualization. **Marie-Christin Langer:** Methodology, Investigation, Conceptualization. **Michael Müller:** Validation, Methodology, Investigation, Conceptualization. **Rudolf Sack:** Validation, Methodology, Investigation, Conceptualization. **Magnus Schlösser:** Writing – review & editing, Validation, Supervision, Resources, Project administration, Methodology, Investigation, Funding acquisition, Conceptualization. **Markus Steidl:** Writing – review & editing, Validation, Supervision, Resources, Project administration, Methodology, Investigation, Funding acquisition, Conceptualization. **Kathrin Valerius:** Writing – review & editing, Validation, Supervision, Resources, Project administration, Methodology, Investigation, Funding acquisition, Conceptualization.

## Declaration of competing interest

The authors declare that they have no known competing financial interests or personal relationships that could have appeared to influence the work reported in this paper.

## Acknowledgments

Funded by the Federal Ministry of Education and Research (BMBF) and the Baden-Württemberg Ministry of Science as part of the Excellence Strategy of the German Federal and State Governments. We further acknowledge support of the Helmholtz Association, Germany and, additionally, the Helmholtz Initiative and Networking Fund under grant agreement no. W2/W3-118. We thank Dr. Beate Bornschein from Tritium Laboratory Karlsruhe (TLK) for fruitful discussions.

[4] Electrons can also undergo scattering inside the source, however, this effect is expected to be small compared to the backscattering on the absorber of the detector.

[5] $KX_{\beta1}$ and $KX_{\beta3}$ are only separated by 8 eV, and cannot be differentiated with the current detector resolution.

[6] From Fig. 7(b): at 30 keV, amplitude deviation from linear behavior is 1.544 keV, giving a non-linearity of 5.15%.

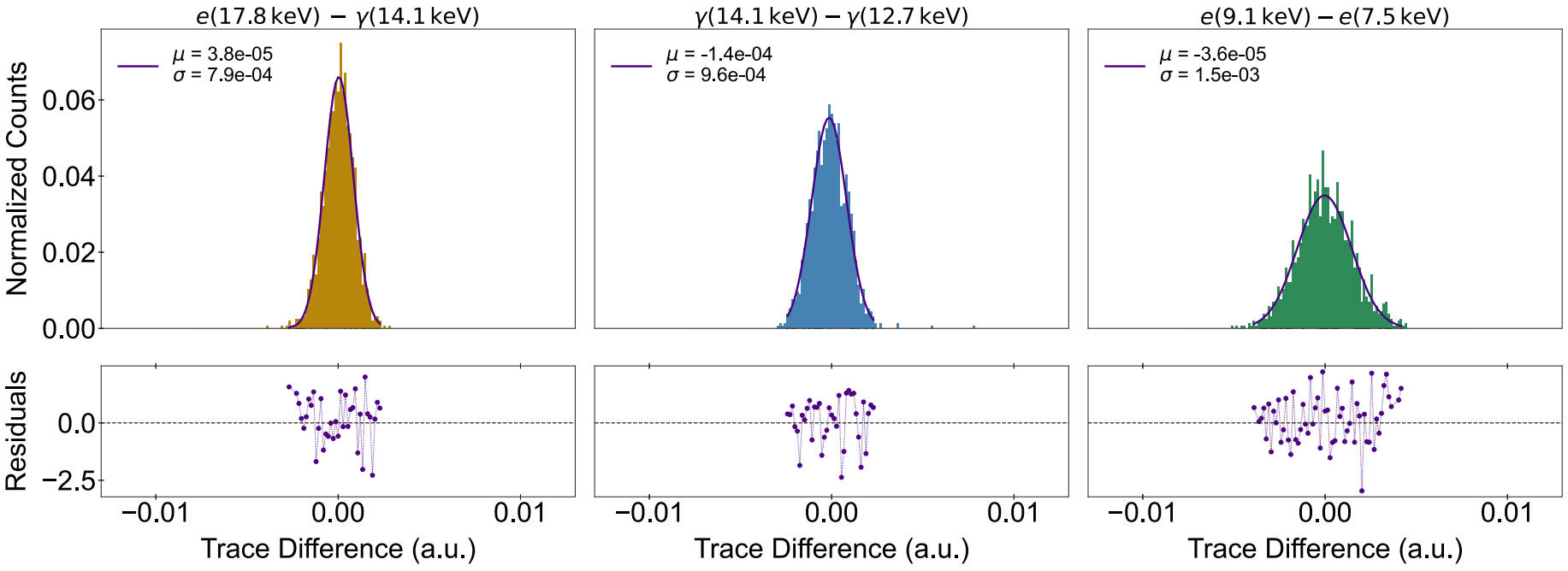

(a) Fits to histograms of signal differences taking into account 0.1 ms around the signal rise

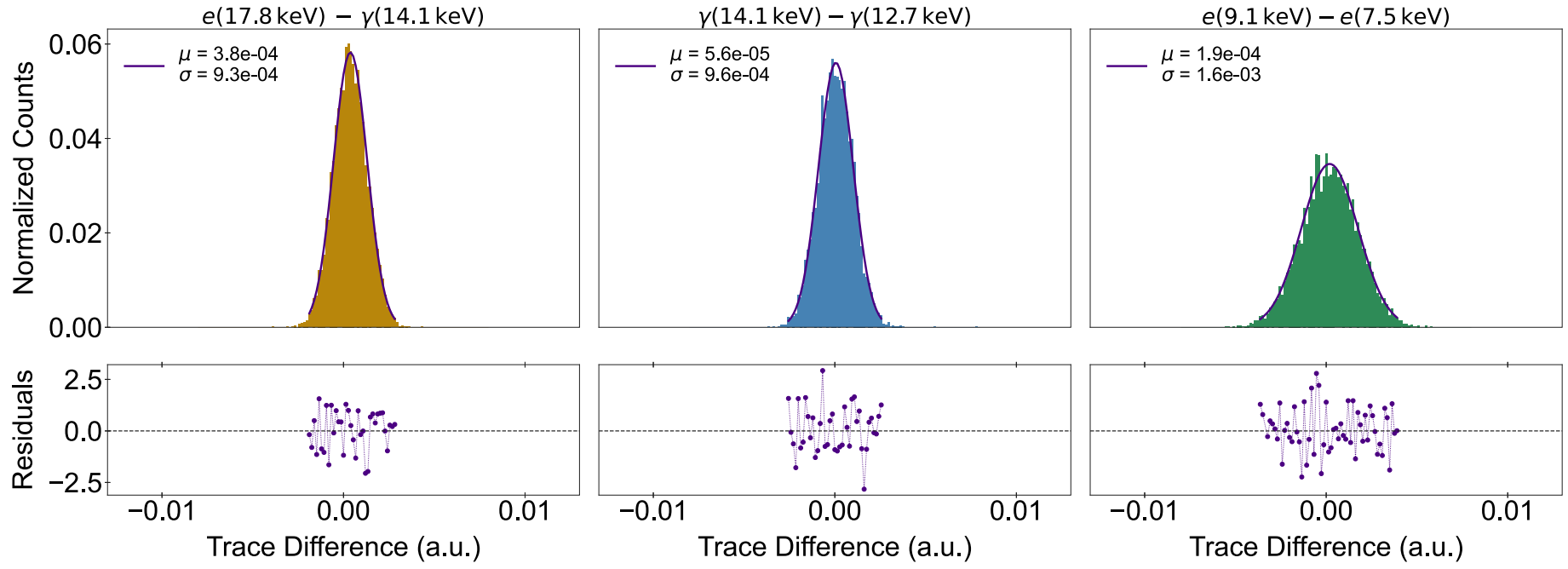

(b) Fits to histograms of signal differences taking into account 0.5 ms around the signal rise

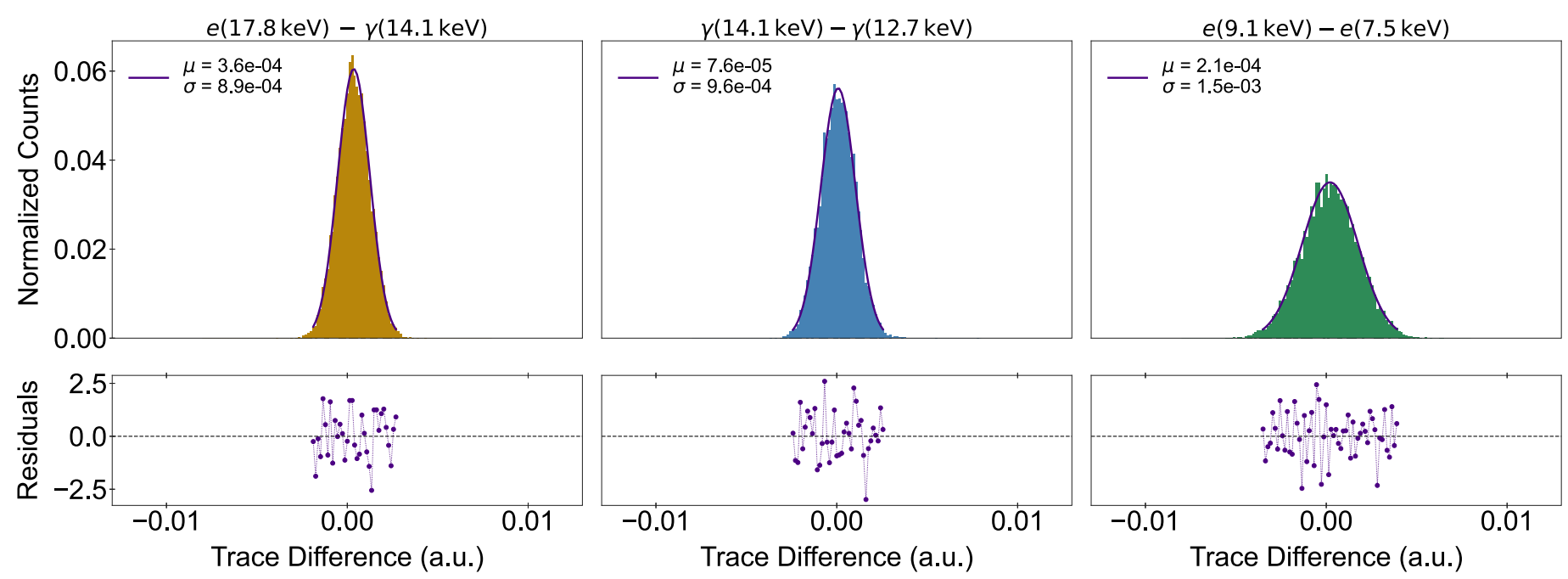

(c) Fits to histograms of signal differences taking into account 1.0 ms around the signal rise

**Fig. A.8.** Fits to the histograms of the differences between individual signal pulses, taking into account signal pulses: (a) within the region of 0.1 ms around the signal rise, (a) within the region of 0.5 ms around the signal rise and (a) within the region of 1.0 ms around the signal rise, as shown in Fig. 5(a). Each histogram is fitted with a Gaussian function, and fit results are reported in the legend. Residuals of the fit are shown below each respective histogram. Plots are shown in linear scale.

## Appendix

In order to confirm that the histograms of pulse differences, shown in Fig. 5, are truly Gaussian, fit have been performed on individual histograms and are shown in Fig. A.8. As can be seen in the figure, all histograms are well described by a Gaussian function, and no visible strictures are present in the residuals. All fit results are summarized in Table 2.

## Data availability

Data will be made available on request.